\documentclass[conference]{IEEEtran}

\usepackage{cite}

\bstctlcite{IEEEexample:BSTcontrol}

\usepackage[pdftex]{graphicx}
\usepackage{epstopdf}
\graphicspath{{./figures/}}
 \DeclareGraphicsExtensions{.pdf,.jpeg,.png,.eps}

\usepackage[cmex10]{amsmath}
\usepackage{verbatimbox}
\usepackage{dsfont,xcolor}

\usepackage[justification=centering]{caption}

\usepackage[protrusion=true,expansion=true]{microtype}
\usepackage{tikz}
\newcommand\copyrighttext{%
	\footnotesize \textcopyright 2016 IEEE. Personal use of this material is permitted.
	Permission from IEEE must be obtained for all other uses, in any current or future
	media, including reprinting/republishing this material for advertising or promotional
	purposes, creating new collective works, for resale or redistribution to servers or
	lists, or reuse of any copyrighted component of this work in other works.
	}
\newcommand\copyrightnotice{%
	\begin{tikzpicture}[remember picture,overlay]
	\node[anchor=south,yshift=10pt] at (current page.south) {\fbox{\parbox{\dimexpr\textwidth-\fboxsep-\fboxrule\relax}{\copyrighttext}}};
	\end{tikzpicture}%
}
\begin{document}
%
\title{Experimental Evaluation of a Reconfigurable Antenna System for Blind Interference Alignment\vspace{-5mm}}
\author{\IEEEauthorblockN{Simon Begashaw, James Chacko, Nikhil Gulati, Danh H. Nguyen, Nagarajan Kandasamy and Kapil R. Dandekar}
	\IEEEauthorblockA{Drexel University, Philadelphia, PA.
		Email: \{sgb42, jjc652, ng54, dnguyen, kandasamy, dandekar\}@drexel.edu}}

\maketitle
\copyrightnotice
\begin{abstract}
In recent years, several experimental studies have come out to validate the theoretical findings of interference alignment (IA), but only a handful of studies have focused on blind interference alignment. Unlike IA and other interference mitigation techniques, blind IA does not require channel state information at the transmitter (CSIT). The key insight is that the transmitter uses the knowledge of channel coherence intervals and receivers utilize reconfigurable antennas to create channel fluctuations exploited by the transmitter. In this work, we present a novel experimental evaluation of a reconfigurable antenna system for achieving blind IA. We present a blind IA technique based on reconfigurable antennas for a 2-user multiple-input single-output (MISO) broadcast channel implemented on a software defined radio platform where each of the receivers is equipped with a reconfigurable antenna. We further compare this blind IA implementation with traditional TDMA scheme for benchmarking purposes. We show that the achievable rates for blind IA can be realized in practice using  measured channels under practical channel conditions. Additionally, the average error vector magnitude and bit error rate (BER) performances are evaluated. 

\end{abstract}


\section{INTRODUCTION}

Due to the increasing demand for high data rates and increasing density of wireless networks, there has been a growing interest in developing advanced interference mitigation techniques such as interference alignment. 
Interference alignment (IA) is a relatively new interference mitigation technique that achieves significant increase in sum rate over traditional orthogonal schemes \cite{Cadambe2008}. 
In a nutshell, IA uses low-complexity precoding to align interfering signals at each receiver into an interference subspace, thereby allowing the intended signal to be decoded in its own signal space. 
Most of the existing approaches to IA place the precoding complexity at the transmitters, with the assumption of perfect, and sometimes global, Tx-Rx channel state information at the transmitter (CSIT). 
This assumption often falls short in practice, as obtaining accurate CSIT requires additional bandwidth and turn-around time that severely impacts the spectral efficiency of the system. 
As a result, the implementation of CSIT-based IA schemes have proven to be challenging.

To address these challenges, a novel technique called blind interference alignment, which does not require CSIT for a certain class of networks, was proposed in~\cite{Jafar2012}. 
Without knowledge of CSIT, the blind IA scheme is able to align interference based on the knowledge of only the channel autocorrelation structures of different users. 
In~\cite{Jafar2010}, blind IA was developed to exploit the staggered block fading nature of the wireless channel for each link to perform alignment. 
To improve upon this, blind IA schemes that leverage reconfigurable antennas to artificially create temporal correlations in the channels have been proposed~\cite{Gou2011},\cite{Wang2010}.

To understand the impact of blind IA techniques on practical wireless networks, it is necessary to evaluate their performance in realistic settings. 
Simulation-based studies often reiterate over a set of simplistic channel models and scenarios, such as spatially uncorrelated channels, perfect timing and frequency synchronization and perfect channel estimation. 
There are only a few experimental evaluations of blind IA schemes in the literature. In~\cite{Miller2012}, a blind IA implementation for the X channel is described and its throughput and BER performance is compared against TDMA. 
Another experimental evaluation that compares a blind IA scheme against Linear Zero Forcing Beamforming (LZBF) is presented in~\cite{Cespedes2015}. 
Both of these works simulate the behavior of reconfigurable antennas using two spatially separated conventional antennas rather than actually employing reconfigurable antennas in their experiments. 
In~\cite{Qian2013}, the performance of blind IA using ESPAR antennas is investigated. The authors show improved performance in terms of ergodic sum rate and BER with the use of the ESPAR antenna. 
Again, this work relies on a simulation of the antenna and not measurements obtained using the ESPAR antenna.
Although relevant, none of these works address practical issues such as short channel coherence time, spatially correlated channels and phase compensations at the receivers.  
For blind IA to be viable for practical communication systems, it is important to experimentally evaluate its performance in realistic channels using compact reconfigurable antennas that can be integrated in mobile devices. 

In this work, we evaluate the performance of a reconfigurable antenna-based blind IA implementation on our multiple-input multiple-output orthogonal frequency division multiplexing (MIMO-OFDM) testbed. 
Reconfigurable antennas have gained significant attention in recent years for both single user systems~\cite{costantine2015reconfigurable,gulati2014learning} as well as multi-user IA based systems~\cite{bahl2012impact,bahl2013reconfigurable}, and to the best of our knowledge, this is the first experimental blind IA work that utilizes reconfigurable antennas instead of simulating their behavior through multiple antennas. 
Reconfigurable antennas for blind IA allow more efficient system design (in terms of cost and space) and performance since a single antenna element on the receiver can generate the required channel fluctuations and removes the requirement for multiple antennas.
The experimental setup consists of one transmitter with two conventional antennas and two users each equipped with a reconfigurable antenna, commonly referred to as a two-user multiple-input single-output broadcast channel (MISO-BC). 
The proposed experiments are based on configurations that are similar to a 802.11-based 2.4 GHz WiFi system. 
Through experimental measurements, we demonstrate that blind IA can indeed be realized in practice via reconfigurable antennas and our reconfigurable antenna-based blind IA implementation significantly outperforms the rate achieved by TDMA. 

The rest of this paper is organized as follows. 
In the next section, we present the system model and blind interference alignment theory. 
Section~\ref{sec:Implementation} describes the testbed, the reconfigurable antenna, the blind IA implementation and measurement setup. 
The results and analysis of the experiments are provided in Section \ref{sec:results}. 
Finally, Section~\ref{sec:conclusion} summarizes the work and provides future direction. 

\section{BACKGROUND}
\label{sec:background}
\subsection{System Model}
\label{sys_model}

To evaluate the performance of reconfigurable antennas in a blind interference alignment implementation, we consider a $K$-user $M\times1$ MISO BC scenario, where the transmitter has M antennas and each of K receivers have a reconfigurable antenna. 
The receivers are able to select one of the preset states of the reconfigurable antenna.

Let $\mathbf{h}^{\left[k\right]}(m)\in \mathds{C}^{2\times1}$ denote the $1 \times M$ channel vector associated with the $m$-th state of user $k$'s reconfigurable antenna. 
As stated in the introduction, blind IA does not require CSI at the transmitter. 
Furthermore, there are no special assumptions made about the channel coherence block structure. However, we do assume that the coherence times are large enough so that the channel vectors stay constant during the symbol extension period required for alignment, commonly referred to as a supersymbol~\cite{Wang2010}. 
During a supersymbol, which is discussed in more detail in section~\ref{(bia_theory)}, the receivers switch between their antenna states in a predetermined pattern. 
Let us denote the state selected by receiver $k$ at time $t$ as $m^{\left[k\right]}(t)$ and the corresponding channel for the user as $ \mathbf{h}^{\left[k\right]}(m^{\left[k\right]}(t))$. 
Suppose signal vector $\mathbf{x}(t) \in \mathds{C}^{M\times1}$ is sent from the transmitter. 
The received signal vector of user $k$ at time $t$ is given by 
\begin{equation}
\label{eq:rx_sig}
y^{\left[k\right]}(t)= \mathbf{h}^{\left[k\right]}(m^{\left[k\right]}(t)) \mathbf{x}(t) + z^{\left[k\right]}(t)
\end{equation}
where $z_{t}^{\left[k\right]}(t)$ represents additive white Gaussian noise with zero mean and unit variance. The channel input is subject to an average power constraint $\mathds{E}\left[ ||x||^{2}\right] \le P$. 
 
\subsection{Blind IA Theory}
\label{(bia_theory)}
The objective of blind interference alignment is to construct signals intended for $K$ different users, such that at each receiver, the signals intended for that receiver remain distinct while the interference (the signals intended for the remaining receivers) cast overlapping shadows. 
This signal construction is achieved without any knowledge of the channel coefficients for each receiver. 
The design of the alignment block or supersymbol structure and the corresponding beamforming strategy is central to the blind interference alignment scheme.
\begin{table}[h]
	\begin{center}
		\renewcommand{\arraystretch}{1.5}
		\begin{tabular}{|c|c|c|c|}
			\hline slot & 1 & 2 & 3 \\
			\hline user 1 & $\mathbf{h}^{\left[1\right]}(1)$ & $\mathbf{h}^{\left[1\right]}(2)$ & $\mathbf{h}^{\left[1\right]}(1)$  \\ 
			\hline user 2 & $\mathbf{h}^{\left[2\right]}(1)$ & $\mathbf{h}^{\left[2\right]}(1)$ & $\mathbf{h}^{\left[2\right]}(2)$  \\ 
			\hline
		\end{tabular}
		\caption{Supersymbol structure for two-user $2 \times 1$ MISO-BC}
		\label{table:supersymbol}
		\vspace*{-20pt}
	\end{center}
\end{table} 
To illustrate the design of the supersymbol structure and the corresponding beamforming strategy, we will focus the subsequent discussion on a $ K = M = 2$ MISO BC case, where there are two users and the transmitter has two antennas. 
Although our analysis focuses on this specific scenario, it has been shown in~\cite{Gou2011} that this scheme can be generalized to the $K$-user $M\times1$ case. 
For the two-user $2 \times 1$ MISO-BC case, the goal is to achieve two degrees of freedom (DoF) for each user over three symbol extensions, for a total of 4/3 DoF. 
This result is accomplished by sending two independent signal streams, each carrying one DoF to each user over a supersymbol. 
As presented in~\cite{Gou2011}, the supersymbol structure for user 1 in a two-user $2 \times 1$ MISO-BC is shown in Table~\ref{table:supersymbol}. 
The table illustrates user 1 using antenna state 1 to receive the signal in the first slot and switching to state 2 in the second slot and returning to state 1 in the third slot. 
User 2 stays in state 1 for the first 2 slots and switches to state 2 for the third slot. 
The alignment block for user 1 is made up of the first 2 slots, while the alignment block for user 2 is made up of slots 1 and 3.\\
\indent At the transmitter, a beamforming strategy has to be designed to leverage the aforementioned supersymbol structure to enable alignment at each of the receivers over the duration of 3 symbol extensions. 
The transmitter has four independent symbols, two for each user. 
The signal vector $\mathbf{u}_{i}^{\left[k\right]} = \left[ u_{1}^{\left[k\right]} , u_{2}^{\left[k\right]} \right]^{T}$
represents the 2 symbols intended for user k. 
To transmit these vectors over 3 symbol extensions, a $6\times2$ beamforming matrix is constructed for each user by stacking three $2 \times 2$ matrices. 
As shown in~(\ref{eq:tx_sig}), the beamforming matrix for each user has a $2\times2$ identity matrix corresponding to the alignment block for each user and a zero matrix in the remaining block. 
With this beamforming matrix, the transmitted signal becomes:
\begin{equation}
\label{eq:tx_sig}
\left[\begin{array}{c}\mathbf{x}(1) \\ \mathbf{x}(2) \\ \mathbf{x}(3)\end{array}\right] =
\left[\begin{array}{c}\mathbf{I}_{2} \\ \mathbf{I}_{2} \\ \mathbf{0}_{2}\end{array}\right]
\left[\begin{array}{c}u_{1}^{\left[1\right]} \\ u_{2}^{\left[1\right]}\end{array}\right] + 
\left[\begin{array}{c}\mathbf{I}_{2} \\ \mathbf{0}_{2} \\ \mathbf{I}_{2}\end{array}\right]
\left[\begin{array}{c}u_{1}^{\left[2\right]} \\ u_{2}^{\left[2\right]}\end{array}\right]
\end{equation}
where $\mathbf{I}_{2}$ and $\mathbf{0}_{2}$ represent a $2\times2$ identity matrix and zero matrix respectively. 
Note that the beamforming vectors do not depend on the values of the channel coefficients. 
With this beamforming strategy, the transmitter is sending two different symbols simultaneously to each user in the first slot. 
During the subsequent time slots, the symbols for each user are transmitted in an orthogonal manner. 
With the supersymbol structure and beamforming matrix discussed above, the received signal at user 1 is given by:
\vspace*{-3mm}
\begin{multline}
\label{eq:rx1_sig}
\left[\begin{array}{c}y^{\left[1\right]}(1) \\ y^{\left[1\right]}(2) \\ y^{\left[1\right]}(3)\end{array}\right] =
\left[\begin{array}{lr}\mathbf{h}_{1}^{\left[1\right]}(1) \\
\mathbf{h}_{2}^{\left[1\right]}(2) \\
\mathbf{0} \end{array}\right]
\left[\begin{array}{c}u^{\left[1\right]}_{1} \\ u^{\left[1\right]}_{2}\end{array}\right] + \\
\left[\begin{array}{lr}\mathbf{h}_{1}^{\left[1\right]}(1) \\
\mathbf{0}  \\
\mathbf{h}_{3}^{\left[1\right]}(1) \end{array}\right]
\left[\begin{array}{c}u^{\left[2\right]}_{1} \\ u^{\left[2\right]}_{2}\end{array}\right] + \left[\begin{array}{c}z^{\left[1\right]}(1) \\ z^{\left[1\right]}(2) \\ z^{\left[1\right]}(3)\end{array}\right]
\end{multline}
where $\mathbf{0}$ is a $1\times2$ zero vector. 
Thus in the three dimensional received signal space of user 1, the interference from the signal intended for user 2 aligns within one dimension along vector $\left[1~0 ~1 \right]^{T}$, while the desired signals, occupy two linearly independent dimensions. 
To obtain its interference free signal, user 1 can use the interference received in the 3rd slot and subtract it from the first slot as shown in~(\ref{eq:ia_decode}):
\begin{multline}
\label{eq:ia_decode}
\left[\begin{array}{c}y^{\left[1\right]}(1)-y^{\left[1\right]}(3) \\ y_{2}^{\left[1\right]}(2)\end{array}\right] =
\left[\begin{array}{lr}h^{\left[1,1\right]}(1) & h^{\left[1,2\right]}(1) \\
h^{\left[1,1\right]}(2) & h^{\left[1,2\right]}(2) \end{array}\right]
\left[\begin{array}{c}u_{1}^{\left[1\right]} \\ u_{2}^{\left[1\right]}\end{array}\right] \\+ 
\left[\begin{array}{c}z^{\left[1\right]}(1) -z^{\left[1\right]}(3)\\ z^{\left[1\right]}(2)\end{array}\right]
\end{multline}
where $h^{\left[i,j\right]}(m)$ represents the coefficient associated with the channel from the $j$-th antenna of the transmitter to receiver $i$ when the $m$-th state of the reconfigurable antenna is selected. 
It is clear from~(\ref{eq:ia_decode}) that user 1 is able to access a full rank channel matrix and therefore can resolve the symbols intended for it and achieve 2 DoF. 
By symmetry, user 2 can follow a similar procedure and cancel out its interference received in the second slot to also achieve 2 DoF, so that a total of 4 DoF are achieved over 3 symbol extensions. 
For the $K$-user $M \times 1$ MISO BC blind IA scheme described here, the authors in \cite{Gou2011} have derived the achievable rate with zero-forcing interference at the receiver as:
\begin{multline}
\label{eq:rate}
R = \sum_{k=1}^{K} \frac{1}{M+K-1} \\ 
 \times \mathds{E}\left[ \log \det (\mathbf{I} +\frac{(K+M-1)P}{M^{2}K}\mathbf{H}^{\left[k\right]}\mathbf{H}^{\left[k\right]\dagger})\right]
\end{multline}
where $\mathbf{H}^{k} = \left[ \frac{1}{\sqrt{K}}\mathbf{h}^{\left[k\right]\dagger}(1)\hspace{1mm}... \frac{1}{\sqrt{K}}\mathbf{h}^{\left[k\right]\dagger}(M-1)\hspace{1mm} \mathbf{h}^{\left[k\right]\dagger}(M)\right]^{\dagger}$ and $P$ is the total transmitted power. 
\section{IMPLEMENTATION OF BLIND IA WITH RECONFIGURABLE ANTENNAS}
\label{sec:Implementation}

In this section, we describe our implementation of the reconfigurable antenna-based blind interference alignment and the experimental testbed we developed to evaluate our implementation. 
We also describe the challenges associated with implementing blind interference alignment on a software defined radio (SDR) platform and the steps we took to address those challenges. 

\subsection{Experiment/Testbed Description}

The experiments were carried out using the WARPLab~\cite{Warplab} framework which facilitates rapid prototyping of physical layer algorithms by combining the signal processing capabilities of MATLAB with the real-time over-the-air (OTA) transmission and reception capabilities of the WARP~\cite{warpProject} SDR.
Within this framework, we digitally process samples on a packet level in MATLAB and transfer them to FPGA buffers via the Ethernet interface for OTA transmission.  
While being extremely efficient for early-stage physical layer prototyping, this canonical WARPLab flow incurs large processing latency and inherently locks the system to a packet-based processing paradigm. 
Once residing in the FPGA buffers, the packet samples cannot undergo any additional signal processing until they are transfered back to the host computer. 
To enable WARPLab to carry out low-latency operations, such as switching antenna states in the middle of packet reception, we augment WARPLab's sample buffer system with custom FPGA signal processing for carrying out time-critical operations. The detailed system modifications are described Section~\ref{implement}.  

Three WARP nodes were used in our experiment, one as a transmitter and two as receivers. 
Each node has two radio boards, allowing us to construct a $2 \times 1$ MISO system, where both radios are used at the transmitter while only one radio was used at the receiver. The experiments were carried out using WiFi channels at 2.4 GHz. 

Using the WARPLab framework, the implemented system has an OFDM based physical layer with a bandwidth of 20 MHz using 64 subcarriers, with 48 subcarriers used for payload. 
OFDM is a suitable choice for our blind interference alignment testbed for a number of reasons. 
One advantage of using OFDM is that it enables us to transform the original frequency selective wireless channel into multiple flat-fading channels. 
This transformation is important for our interference alignment implementation since it allows for alignment on a subcarrier basis. 
Additionally, the duration of an OFDM symbol is considerably larger than the duration of symbols in many non-OFDM systems. 
This larger symbol duration helps to improve timing error tolerance on the otherwise stringent requirement of symbol-level synchronization imposed by interference alignment. 
\begin{figure}[!t]
	\centering
	\includegraphics[width=3.5in]{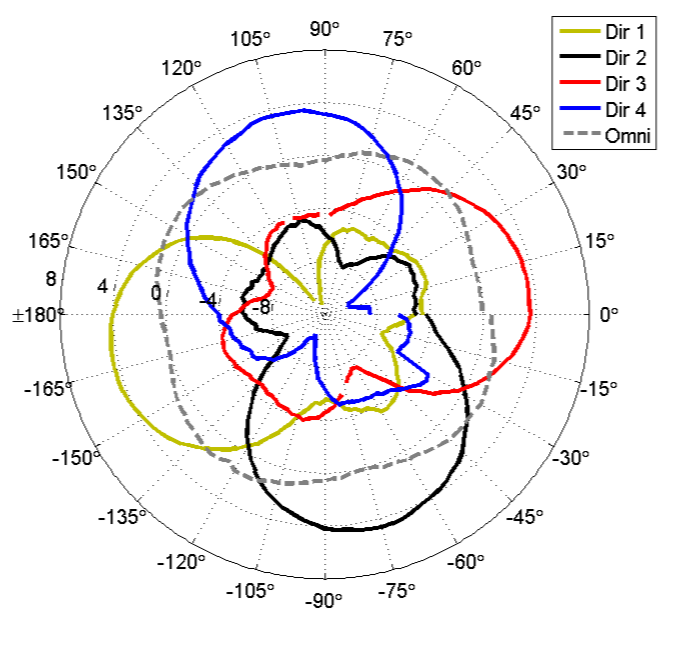}
	\caption{Directional and omni-directional radiation patterns of the Reconfigurable Alford Loop Antenna}
	\label{fig:pattern}
	\vspace*{-1\baselineskip}
\end{figure}
\subsection{Reconfigurable Antenna}
\label{rec_ant}
In this work, we employ the Reconfigurable Alford Loop Antenna~\cite{Patron2014}, a planar reconfigurable antenna with integrated control circuitry designed by the Drexel University Wireless Systems Laboratory (DWSL). 
This pattern reconfigurable antenna is composed of four~$90^{\circ}$ microstrip elements placed symmetrically on a substrate and connected to a central feed port. 
The elements can be individually switched on and off with the use of PIN diodes. 
When all the elements are turned on, the resulting radiation pattern of the antenna is omni-directional in the azimuth plane. 
Alternatively, each element can be individually turned on to generate four directional beams with~$90^{\circ}$ spacing. 
The four directional and the omni-directional measured patterns are displayed in Fig.~\ref{fig:pattern}. 
It is also possible to use different combinations of elements to generate additional directional or bidirectional beam patterns. 
One of the reasons this antenna was chosen for our blind IA implementation was because of the integrated control circuitry, which makes it convenient for deployment with SDR. 
Common GPIO voltages found on SDR platforms such as 3.3 V or 5V can be used to select antenna states. Additionally, the antenna has a compact design (65 mm $\times$ 65 mm) that makes it practical for integration into mobile devices for the use of blind IA and other applications that leverage radiation pattern diversity. 
Each of the receiver WARP nodes in our experimental setup is equipped with this antenna. 
The user GPIO pins on the WARP board are used to control the antenna states. 

\subsection{Implementation}
\label{implement}
An important requirement of any blind IA system is that the channel coherence time must be longer than the supersymbol time over which alignment occurs. 
An early version of our experimental blind IA testbed was developed using the standard WARPLab experimental flow, in which an entire packet buffer was sent at a time and interference alignment was performed on a packet basis. 
Following this packet-based processing paradigm, we constructed the supersymbol out of three packets. 
The first packet would contain both the signal intended for user~1 and the signal intended for user~2 while the second packet would only contain the data stream for user 1 and the third packet would contain only contain the data stream for user~2. 
The receiver would select the antenna state to receive each of the~3 packets that constitute an alignment block. 
Only after receiving the~3 packets could the OFDM demodulation, interference suppression, and symbol detection take place. 
With this kind of implementation, the channel must stay constant over the time required to transmit and receive three packets for alignment to be successful. 
Channel sounding measurements were carried out to determine whether the channel coherence time for the measurement environment was larger than the time required for the transmission and reception of~3 packets using the WARPLab framework. 
The OTA measurements showed that this channel condition was not always satisfied. Using the standard WARPLab framework, the time required to transmit and receive three packets, while also making changes the receiver's antenna states between packet transmissions is approximately one second. Most of this delay comes from the time required to download and upload samples from the host computer to the FPGA buffers and from the Ethernet-interfaced functions that select the antenna state for receiving each packet. Over this one second period, the variation in the magnitude of the channel coefficients was less than 2 dB. However, there were significant variations, ranging between $10^{\circ}-15^{\circ}$, in the phase of channel coefficients. 
The presence of phase offsets leads to interference leakage into the signal subspace since it cannot be canceled out effectively using the method described in section \ref{(bia_theory)}.

To address this issue, the testbed was modified to enable implementation of interference alignment at the OFDM symbol level rather than at the packet level. 
This implementation significantly reduces the constraint on the channel coherence time. Since the time required to transmit individual OFDM symbols is much smaller than the time required to transmit packets, the assumption that the channel coherence time is longer than the duration of the alignment block becomes valid. 
In fact, it is possible to group multiple OFDM symbols to transmit within one slot of the supersymbol. 
This was verified with measurements. 
In order to implement symbol-level alignment, certain time critical tasks such as packet detection and antenna state selection were moved from WARPLab to the FPGA. 
In the WARPLab framework, antenna state is selected before packet transmission while packet detection is carried out in MATLAB using the data captured into the receive buffers of the WARP nodes. 
For symbol-level alignment, antenna switching needs to occur in real-time at the OFDM symbol level and the MATLAB Ethernet-interfaced functions were not fast enough for switching antenna configurations within the duration of a supersymbol. 
In order to overcome this timing constraint, we implemented a packet detector on the on-board FPGA and provided the logic to have the FPGA select the states of the reconfigurable antenna upon detection of a packet.

The cross correlation based packet detector for the purpose of this project was imported from the Drexel Software Defined Communication (SDC)~\cite{shishkin2011sdc} testbed. 
The SDC packet detector design was chosen for its register-interface based control of detection parameters and its ease in scaling to perform a 256 point correlation that we required. 
Once the preamble is detected, the precise antenna switching time is a known offset from the beginning of the frame. 
With these changes to the hardware, we were capable of configuring both the duration of the supersymbol slots and the antenna states used to receive the signals in each of those slots, allowing flexible experiments that required real time symbol-level antenna state switching

In our symbol-level blind IA implementation, the transmitter sends $N$ symbols in each of the three slots of the supersymbol. Let $\mathbf{u}_{i}^{\left[k\right]}$ represent the $1\times N$ symbol vector intended for user $k$ that is transmitted by antenna $i$. 
The signal vectors $\mathbf{u}_{1}^{\left[1\right]}+\mathbf{u}_{2}^{\left[2\right]}$, $\mathbf{u}_{2}^{\left[1\right]}+\mathbf{u}_{2}^{\left[2\right]}$ are sent from the transmitter's two antennas during the first slot of the supersymbol. 
Consequently, $\mathbf{u}_{1}^{\left[1\right]}$,$\mathbf{u}_{2}^{\left[1\right]}$ and $\mathbf{u}_{1}^{\left[2\right]}$, $\mathbf{u}_{2}^{\left[2\right]}$ are transmitted in the second and third slot respectively. 
By default, both user 1 and user 2 have the omni-directional state of their antenna selected to facilitate packet detection. 
Upon packet detection, user 1 will select its directional state 1 to receive the first $N$ OFDM symbols of the payload that constitute the first slot. 
Then, it switches to the second  state to receive the next $N$ symbols that constitute the second slot and returns to state 1 to receive the remaining $N$ symbols of the packet. The receiver then performs OFDM demodulation, interference cancellation, and aligned symbol detection. User 2, on the other hand, receives the first 2 slots in antenna state 1 and switches to state~2 for the third slot. 
To benchmark our blind IA implementation, we also ran the experiments in TDMA mode where time slots were assigned for each user and the data for each user was transmitted in orthogonal time slots.
Each of the~2 users in this TDMA mode access an interference free SISO channel for half the number of time slots. The duration of each slot in the TDMA scheme is the same as the duration of each slot of the supersymbol in the blind IA implementation. 

\section{RESULTS}
\label{sec:results}
In this section we present the measurement results from the experiment setup described in the previous section. 
We validate that blind IA technique leveraging reconfigurable antennas can indeed be achieved in practical channel conditions and further compare our blind IA implementation with TDMA using three evaluation metrics: sum rates, average error vector magnitude squared, and bit error rate. 
\begin{figure}[!t]
	\centering
	\includegraphics[width=3.3in]{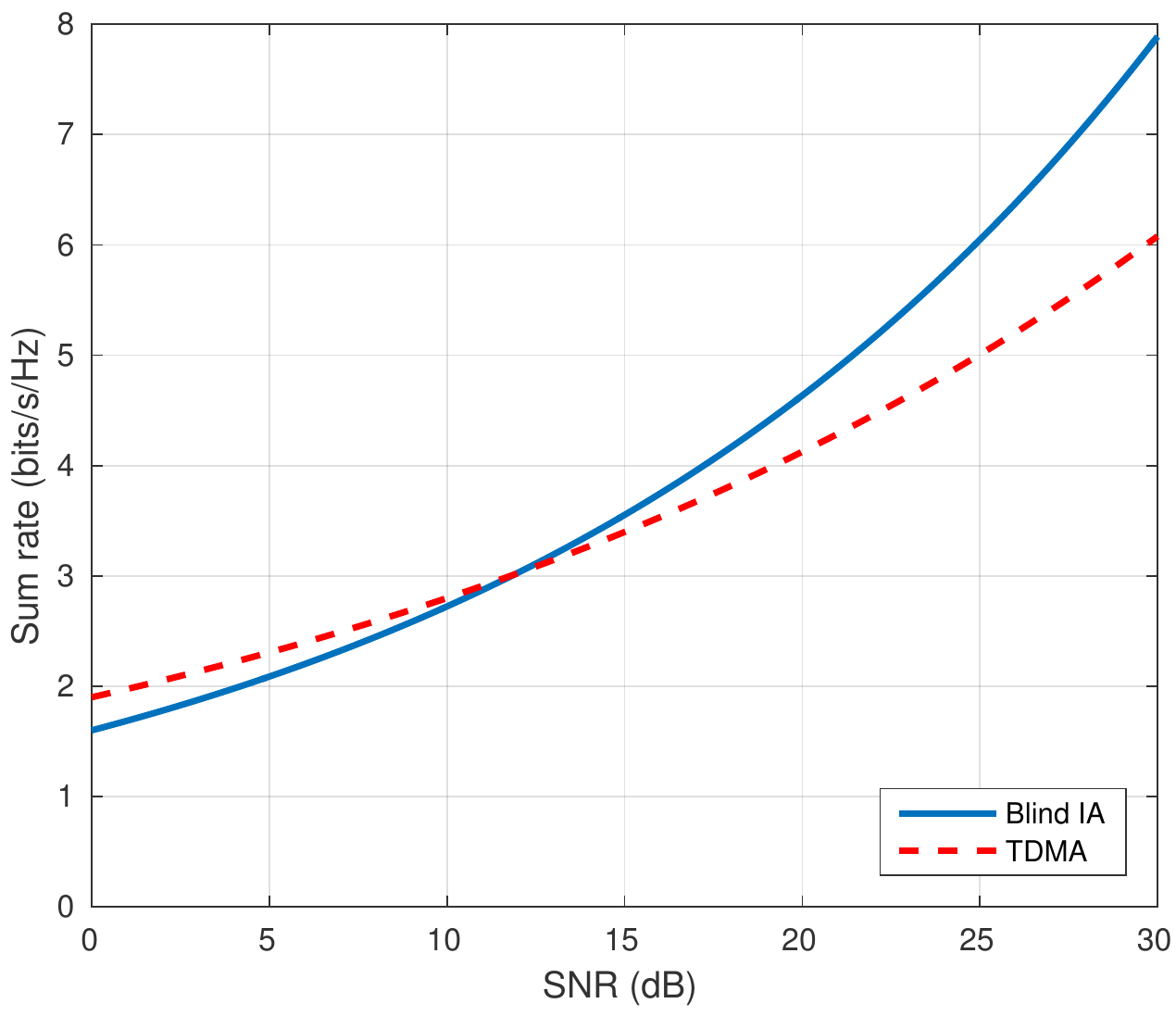}
	\caption{Sum rates performance of blind IA vs TDMA}
	\label{fig:sumrate}
	\vspace*{-2.5mm}
\end{figure}
\subsection{Sum Rate Performance}
We begin our evaluation by looking at the sum rate performance of blind IA and TDMA. 
The sum rates are estimated from measured data. Fig. \ref{fig:sumrate} shows that blind IA clearly outperforms TDMA for high SNR regions. 
Specifically we observe that for SNR values greater than~10 dB, the capacity of blind IA grows at a much faster rate than TDMA. 
For low SNR, TDMA has better rate performance. At SNR of~30 dB, blind IA achieves a rate of approximately~8 bits/s/Hz while the TDMA rate is just above~6 bit/s/Hz. 
This ratio is approaching the theoretical rate gain of~4/3 achieved by IA over orthogonal schemes such as TDMA for the $2$-user $2\times 1$ MISO-BC case. 

\subsection{Average Error Vector Magnitude Performance}
The second result presented is the average error vector magnitude squared (AEVMS) performance of the implementation. 
The error vector is defined as the difference between the received constellation points and the true constellation points and AEVMS, for a normalized constellation is given by 
\begin{equation}
AEVMS = \mathds{E}\left[ |s\left[i\right]-\hat{s}\left[i\right]|^{2}\right]
\end{equation}
where $s\left[i\right]$ and $\hat{s}\left[i\right]$ represent the $i$-th received and ideal symbols respectively. 
AEVMS is a suitable metric for evaluation of hardware experiments for a few reasons. 
First, it can be easily measured since it is computed at the input of the demodulator. 
More importantly, AEVMS captures both the channel induced imperfections such as channel estimation errors and implementation-induced imperfections such as timing errors as well as hardware distortions. 
Furthermore, the authors in~\cite{Duarte2010} and~\cite{Shafik2006} have shown that $1/AEVMS$ can be used to approximate signal to interference and noise ratio (SINR). 
This method of SINR estimation, which we will refer to as post processing SINR (PP-SINR), is more suitable for hardware evaluation than estimations based on energy per symbol to noise ratio ($E_{s}/N_{0}$) or energy per bit to noise ratio ($E_{b}/N_{0}$) which are difficult to measure accurately due to the non-linear nature of noise in hardware~\cite{Duarte2010}.
\begin{figure}[!t]
	\centering
	\includegraphics[width=3.6in]{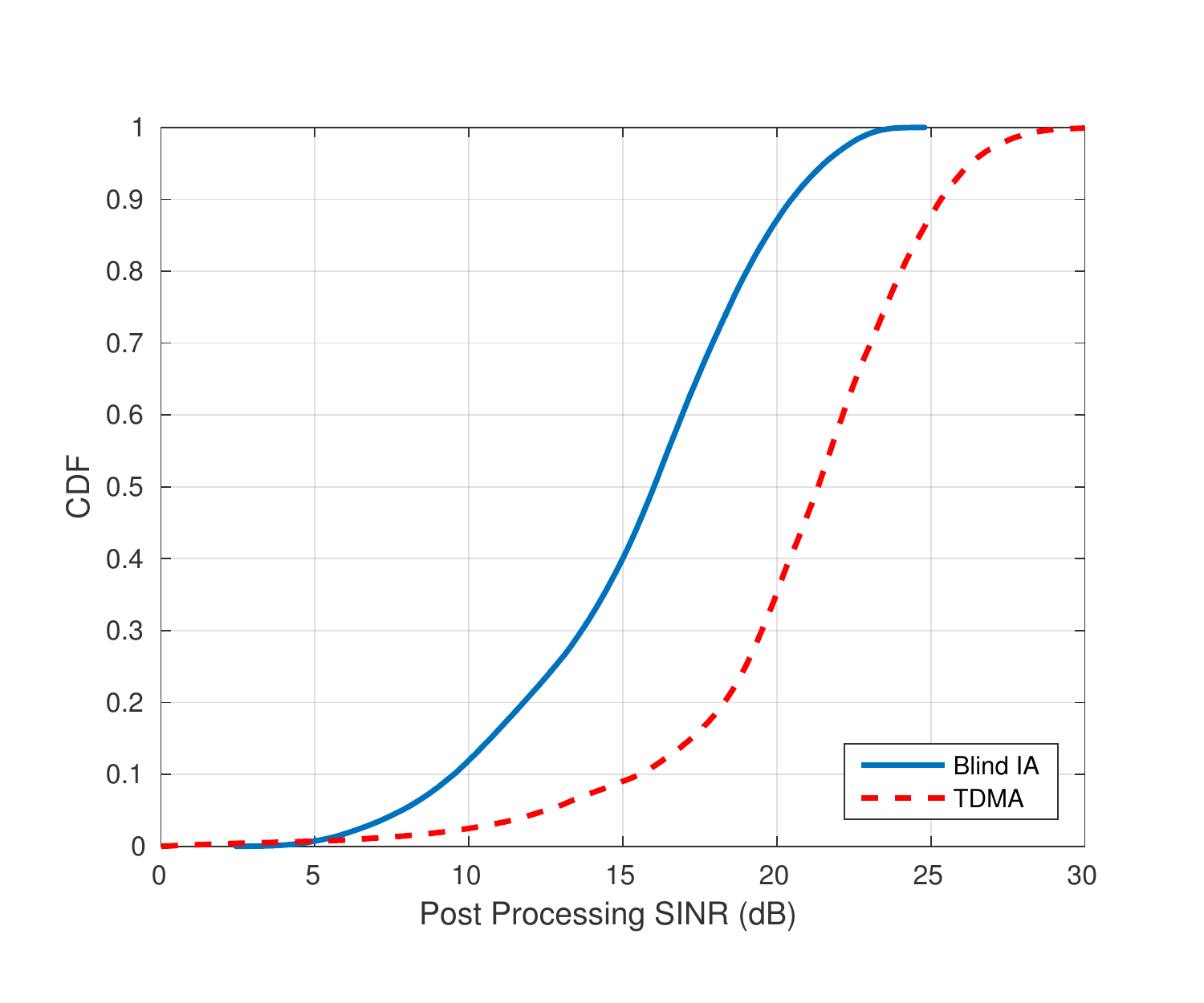}
	\caption{CDF of PP-SINR}
	\label{fig:cdf}
	\vspace*{-3mm}
\end{figure}
We present, in Fig.~\ref{fig:cdf}, the cumulative distribution function (CDF) of the PP-SINR of blind IA and TDMA measurements. 
We expect TDMA to have better PP-SINR performance because the transmitter uses different time slots to send data to each receiver and therefore the transmission is interference free. 
Blind IA involves simultaneous transmissions of data to both users and inherently has interference, which leads to lower PP-SINR performance. 
We can see from Fig.~\ref{fig:cdf} that PP-SINR degradation in blind IA compared to TDMA is less than~5 dB over the entire distribution. 
\subsection {Bit Error Rate Performance}
\begin{figure}[!t]
	\centering
	\includegraphics[width=3.5in]{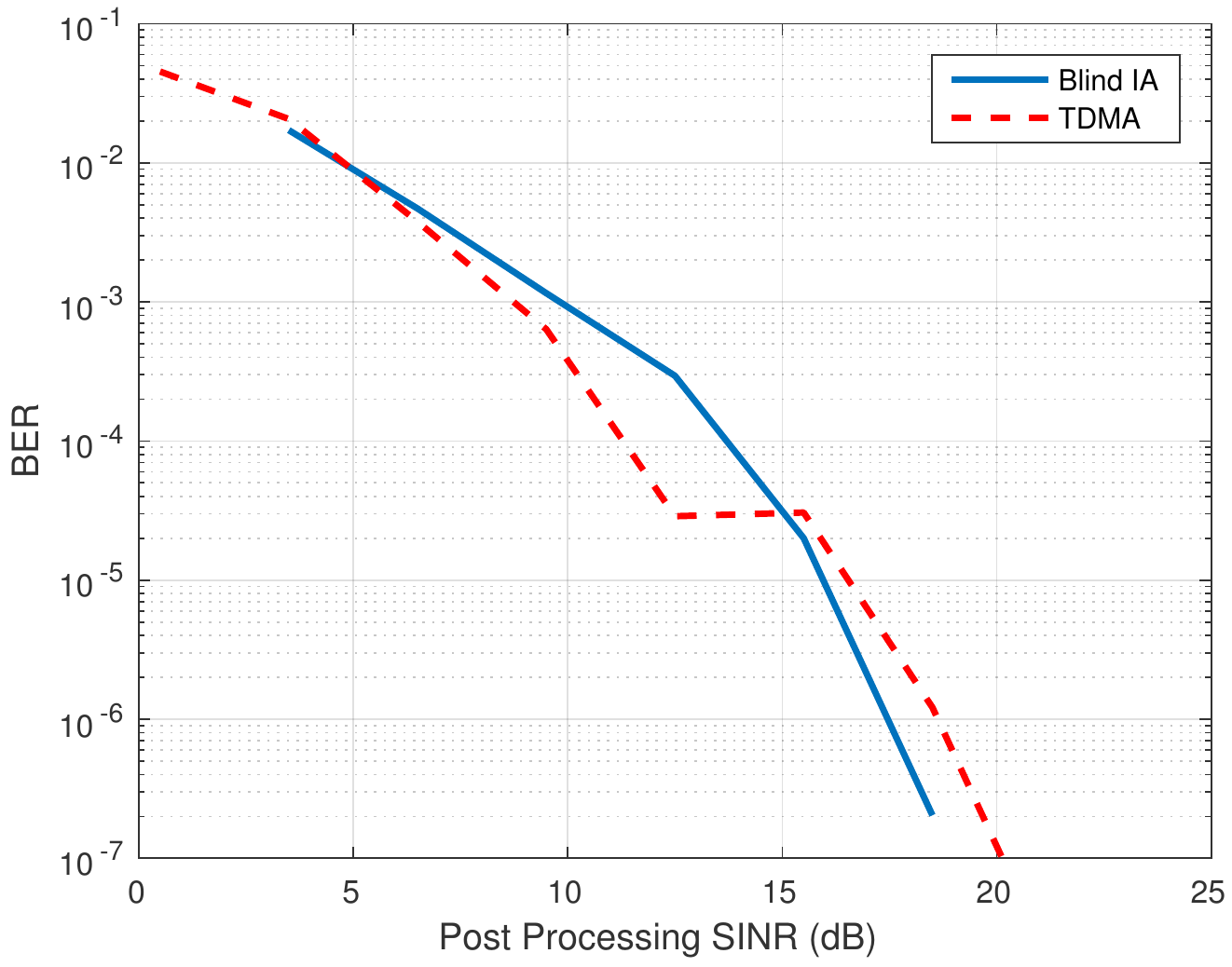}
	\caption{BER performance of Blind IA and TDMA}
	\label{fig:ber}
	
\end{figure}

Finally, we present the BER performance for the two systems in Fig.~\ref{fig:ber}. 
The key take away from this plot is that the BER performance for a given PP-SINR is very similar in both systems. 
This indicates that the extra physical layer processing such as interference suppression and MIMO channel equalization in our blind IA implementation does not degrade the BER performance in relation to TDMA, which has an interference-free SISO channel and has a lower rate than blind IA. 
\section{CONCLUSION}
\label{sec:conclusion}
In this paper, we presented an experimental study of a blind interference alignment scheme that employs a pattern reconfigurable antenna. Unlike other interference mitigation techniques such as beamforming or IA, our reconfigurable antenna-based blind IA implementation does not require CSIT.  
Using our MIMO-OFDM testbed and the Reconfigurable Alford Loop Antenna, we validated the practicality of realizing blind IA with a reconfigurable antenna. Furthermore, we studied the performance of our implementation and how it compares to TDMA. 
Our measurement results show that the  implementation with this antenna achieves significant gain in sum rates compared to TDMA. Due to the inherent interference of blind IA, our implementation incurs 5 dB degradation in terms of PP-SINR. However, for a given PP-SINR, both blind IA and TDMA have similar performance.
Because the Reconfigurable Alford Loop antenna used in this work has several radiation patterns to choose from, a natural extension of our work is the study of optimal antenna pattern selection for blind IA. 

\section*{ACKNOWLEDGMENTS} 
The research presented in this paper was based upon work supported by the
National Science Foundation under Grant No. CNS-1422964.

\bibliographystyle{myIEEEtran}
\bibliography{./bib/BIA_refs}

\begin{thebibliography}{10}
\providecommand{\url}[1]{#1}
\csname url@samestyle\endcsname
\providecommand{\newblock}{\relax}
\providecommand{\bibinfo}[2]{#2}
\providecommand{\BIBentrySTDinterwordspacing}{\spaceskip=0pt\relax}
\providecommand{\BIBentryALTinterwordstretchfactor}{4}
\providecommand{\BIBentryALTinterwordspacing}{\spaceskip=\fontdimen2\font plus
\BIBentryALTinterwordstretchfactor\fontdimen3\font minus
  \fontdimen4\font\relax}
\providecommand{\BIBforeignlanguage}[2]{{%
\expandafter\ifx\csname l@#1\endcsname\relax
\typeout{** WARNING: IEEEtran.bst: No hyphenation pattern has been}%
\typeout{** loaded for the language `#1'. Using the pattern for}%
\typeout{** the default language instead.}%
\else
\language=\csname l@#1\endcsname
\fi
#2}}
\providecommand{\BIBdecl}{\relax}
\BIBdecl

\bibitem{Cadambe2008}
V.~R. Cadambe and S.~A. Jafar, ``{Interference alignment and degrees of freedom
  of the K-user interference channel},'' \emph{IEEE Transactions on Information
  Theory}, vol.~54, no.~8, pp. 3425--3441, 2008.

\bibitem{Jafar2012}
\BIBentryALTinterwordspacing
S.~A. Jafar, ``{Blind Interference Alignment},'' \emph{IEEE Journal of Selected
  Topics in Signal Processing}, vol.~6, no.~3, pp. 216--227, Jun. 2012.
\BIBentrySTDinterwordspacing

\bibitem{Jafar2010}
\BIBentryALTinterwordspacing
------, ``{Exploiting Channel Correlations - Simple Interference Alignment
  Schemes with No CSIT},'' in \emph{IEEE Global Telecommunications Conference
  GLOBECOM 2010}.\hskip 1em plus 0.5em minus 0.4em\relax IEEE, Dec. 2010.
\BIBentrySTDinterwordspacing

\bibitem{Gou2011}
T.~Gou, C.~Wang, and S.~A. Jafar, ``{Aiming perfectly in the dark-blind
  interference alignment through staggered antenna switching},'' \emph{IEEE
  Transactions on Signal Processing}, vol.~59, no.~6, pp. 2734--2744, 2011.

\bibitem{Wang2010}
\BIBentryALTinterwordspacing
C.~Wang, T.~Gou, and S.~A. Jafar, ``{Interference alignment through staggered
  antenna switching for MIMO BC with no CSIT},'' in \emph{2010 Conference
  Record of the Forty Fourth Asilomar Conference on Signals, Systems and
  Computers}, no.~1.\hskip 1em plus 0.5em minus 0.4em\relax IEEE, Nov. 2010,
  pp. 2081--2085.
\BIBentrySTDinterwordspacing

\bibitem{Miller2012}
\BIBentryALTinterwordspacing
K.~Miller, A.~Sanne, K.~Srinivasan, and S.~Vishwanath, ``{Enabling real-time
  interference alignment},'' in \emph{Proceedings of the thirteenth ACM
  international symposium on Mobile Ad Hoc Networking and Computing - MobiHoc
  '12}.\hskip 1em plus 0.5em minus 0.4em\relax ACM Press, 2012, p.~55.
\BIBentrySTDinterwordspacing

\bibitem{Cespedes2015}
\BIBentryALTinterwordspacing
M.~M. Cespedes, M.~S. Fernandez, and A.~G. Armada, ``{Experimental Evaluation
  of Blind Interference Alignment},'' in \emph{2015 IEEE 81st Vehicular
  Technology Conference (VTC Spring)}.\hskip 1em plus 0.5em minus 0.4em\relax
  IEEE, May 2015, pp. 1--5.
\BIBentrySTDinterwordspacing

\bibitem{Qian2013}
\BIBentryALTinterwordspacing
R.~Qian and M.~Sellathurai, ``{Performance of the blind interference alignment
  using ESPAR antennas},'' in \emph{2013 IEEE International Conference on
  Communications (ICC)}.\hskip 1em plus 0.5em minus 0.4em\relax IEEE, Jun.
  2013, pp. 4885--4889.
\BIBentrySTDinterwordspacing

\bibitem{costantine2015reconfigurable}
J.~Costantine, Y.~Tawk, S.~E. Barbin, and C.~G. Christodoulou, ``Reconfigurable
  antennas: Design and applications,'' \emph{Proceedings of the IEEE}, vol.
  103, no.~3, pp. 424--437, 2015.

\bibitem{gulati2014learning}
N.~Gulati and K.~R. Dandekar, ``Learning state selection for reconfigurable
  antennas: A multi-armed bandit approach,'' \emph{Antennas and Propagation,
  IEEE Transactions on}, vol.~62, no.~3, pp. 1027--1038, 2014.

\bibitem{bahl2012impact}
R.~Bahl, N.~Gulati, K.~R. Dandekar, and D.~Jaggard, ``Impact of pattern
  reconfigurable antennas on interference alignment over measured channels,''
  in \emph{Globecom Workshops (GC Wkshps), 2012 IEEE}.\hskip 1em plus 0.5em
  minus 0.4em\relax IEEE, 2012, pp. 557--562.

\bibitem{bahl2013reconfigurable}
R.~Bahl, N.~Gulati, K.~R. Dandekar, and D.~L. Jaggard, ``Reconfigurable
  antennas for performance enhancement of interference networks employing
  interference alignment,'' Jun.~19 2013, uS Patent App. 14/408,807.

\bibitem{Warplab}
"WARPLab".[Online]. Available:https://warpproject.org/trac/wiki/WARPLab.

\bibitem{warpProject}
"WARP Project".[Online]. Available: {http://warpproject.org}.

\bibitem{Patron2014}
\BIBentryALTinterwordspacing
D.~Patron and K.~R. Dandekar, ``{Planar reconfigurable antenna with integrated
  switching control circuitry},'' in \emph{The 8th European Conference on
  Antennas and Propagation (EuCAP 2014)}.\hskip 1em plus 0.5em minus
  0.4em\relax IEEE, Apr., pp. 2737--2740.
\BIBentrySTDinterwordspacing

\bibitem{shishkin2011sdc}
B.~Shishkin, D.~Pfeil, D.~Nguyen, K.~Wanuga, J.~Chacko, J.~Johnson,
  N.~Kandasamy, T.~P. Kurzweg, and K.~R. Dandekar, ``{SDC} testbed: Software
  defined communications testbed for wireless radio and optical networking,''
  in \emph{{Modeling and Optimization in Mobile, Ad Hoc and Wireless Networks
  (WiOpt), 2011 International Symposium on}}.\hskip 1em plus 0.5em minus
  0.4em\relax IEEE, 2011, pp. 300--306.

\bibitem{Duarte2010}
M.~Duarte, A.~Sabharwal, C.~Dick, and R.~Rao, ``{Beamforming in MISO systems:
  Empirical results and EVM-based analysis},'' \emph{IEEE Transactions on
  Wireless Communications}, vol.~9, no.~10, pp. 3214--3225, 2010.

\bibitem{Shafik2006}
\BIBentryALTinterwordspacing
R.~A. Shafik, M.~S. Rahman, and A.~R. Islam, ``{On the Extended Relationships
  Among EVM, BER and SNR as Performance Metrics},'' in \emph{2006 International
  Conference on Electrical and Computer Engineering}.\hskip 1em plus 0.5em
  minus 0.4em\relax IEEE, December 2006, pp. 408--411.
\BIBentrySTDinterwordspacing

\end{thebibliography}

\end{document}